\documentstyle[11pt,newpasp,twoside,epsfig]{article}

\begin{document}
\newcommand{\strp}{\mcol{1}{c|}{-} }
\newcommand{\mcol}[3]{\multicolumn{#1}{#2}{#3} }
\newcommand{\struut}{\rule[-2ex]{0ex}{5.2ex}}
\newcommand{\struutup}{\rule{0ex}{3.2ex}}
\newcommand{\struutdown}{\rule[-2ex]{0ex}{2ex}}

\title{$\delta$ Scuti Stars in Stellar Systems: the interest of HD 220392 and HD 220391\footnotemark[1]
}
\author{P. Lampens and M. Van Camp}
\affil{Koninklijke Sterrenwacht van Belgi\"{e}, Ringlaan 3, B-1180 Brussel, Belgium}

\begin{abstract}
The very wide double star CCDM~23239-5349 is an interesting study case
of a pulsating star within a common origin pair or wide binary. {\it
Both} stars are located in the $\delta$ Scuti instability strip. Based
on data obtained at La Silla (Chile), at the Swiss 0.7m and the ESO 0.5m
telescopes, we found two periodicities of about 4.7 and 5.5 cycles per day
(cpd) with amplitudes of 0.014 and 0.011 mag, respectively, for HD~220392,
the brightest member of the system. The most dominant periodicity was also
detected in the Hipparcos Epoch Photometry data. A similar period search
on the (smaller) dataset obtained for the 1 mag fainter B-component,
HD~220391, however shows no periodic behaviour with an amplitude
significantly above the noise level of the data (about 0.006~mag).
\end{abstract}


\section{Introduction}

\footnotetext[1]{ Based on observations done at La Silla (ESO, Chile) and on data
 obtained by the Hipparcos astrometry satellite.}

The double star CCDM~23239-5349 is a wide visual system consisting of two bright
stars (with $\Delta${\rm m} $\simeq$ 1 mag and an angular separation of 26.5 arcsec)
having similar proper motions as well as compatible parallaxes. This classifies
it as a wide binary (see Sect.4.1).

Regular short-period light variations on a time scale of $\simeq$ 5 hr
have been detected for the brightest component of the system (Lampens~1992). 
The detailed investigation of the difference in variability and physical properties 
between two components of a stellar pair is particularly interesting when 
both companions are located in the same area of the colour-magnitude diagram: in this
case both stars are situated in the $\delta$ Scuti instability strip. The aim of this 
study is to search for clues to understand which factors determine the pulsation 
characteristics such as modes and amplitudes among $\delta$ Scuti stars.
A more extensive discussion will appear elsewhere (Lampens, Van Camp, \& Sinachopoulos~2000).

\section{Observations and reduction}

The photometric data have been gathered during three campaigns
at La Silla: two campaigns performed at the 0.7m Swiss telescope
(P. Lampens) plus one at the 0.5m ESO telescope (D. Sinachopoulos).
In Table~\ref{photo} we display the amount of data taken for the targets
HD~220392 (396 data) and HD~220391 (245 data).  All Geneva
data are absolute measurements in the filters UBVB$_{1}$B$_{2}$V$_{1}$G of
the Geneva Photometric System acquired with standard star measurements
and obtained via a centralized reduction method (Rufener~1988).
This centralized processing has not been applied to the ESO data taken
in the UBV photometric system. The reduction of the October data implied
using a check-star HD~220729 (V=5.52, sp.~type F4V) whose measurements
were interpolated between the two other ones. We have verified the
constancy of this star ($H_{p}=5.6197\,$mag,$\sigma _{H_{p}}=0.0005\,$mag)
in the Hipparcos Catalogue (ESA~1997) and we have fitted a 5th degree
polynomial to the check-star observations for each night separately. 
This polynomial was then subtracted from the data of both programme stars in
order to  suppress as well as possible common variations. The ESO data
are thus interpreted as differential measurements only.

In addition we made use of the data in the Hipparcos Epoch Photometry Catalogue (ESA~1997). 

\begin{table}[]
\caption{Photometric data available for HD~220392 and HD~220391
\label{photo} } 
\begin{center}
{\small
\begin{tabular}{lllccl}
\hline
Identifier & Instrument & Epoch \parbox[b]{0.1cm}{\mbox{} \vspace{1.2em}} & Number & Number & Time base \\ 
&  & \parbox[t]{0.1cm}{\mbox{} \vspace{0.5em}} & of data & of nights &{\it (days) } \\ \hline
HD 220392 & 0.7mGEN & June 90-Sept. 91 & 124 & 18 & 464 \struutup \\ 
 & 0.7m+0.5mESO & June 90-Oct. 92 & 396 &  21 & 866 \\ 
 & Hipparcos & Nov. 89-Mar. 93 & 176 & - & {\it 41 months} \\ 
HD 220391 & 0.7mGEN & Sept. 1991 & 98 & 11 & 19.1 \struutup \\
 & 0.7m+0.5mESO & Sept. 91-Oct. 92 & 245 &  14 & 416 \\ 
 & Hipparcos & Nov. 89-Mar. 93 & 172 &  - & {\it 41 months} 
\\[0.2em]
\hline 
\end{tabular} }
\end{center}
\end{table}

\section{Period analyses}

\subsection{HD~220392}

We combined the 3 nights of ESO V~data with the Geneva V~magnitudes 
into a total of 396 V~data with a time base of 866 days by adjusting the mean
value of the ESO (differential) data to the mean Geneva V magnitude. We made use
of PERIOD98 (Sperl 1998) for the frequency analyses and tested different combinations 
with the abovementioned dataset which confirm the results obtained with the Geneva data only. 
The results of these analyses are displayed in Tables~2b (Fourier fit) and~3 (frequency search) 
(Lampens, Van Camp, \& Sinachopoulos 1999). Two alternative solutions are mentioned. After prewhitening 
for 4.67 cpd, a second frequency of 5.52 cpd was chosen as the next most dominant frequency 
because of a slightly higher reduction of the residual standard deviation of the largest dataset.
The mean light curves presented in Figure~1 have amplitudes of 0.014 and 0.011~mag 
respectively: left is a plot of all the data against a frequency of 4.67439 cpd (after having 
taken the 5.52 cpd variation into account) while right shows the same but against a 
frequency of 5.52234 cpd. The residual dispersion after two prewhitenings amounts to 0.006 mag, 
which is of the order of the noise level in the data.

\begin{figure}[]
\plotfiddle{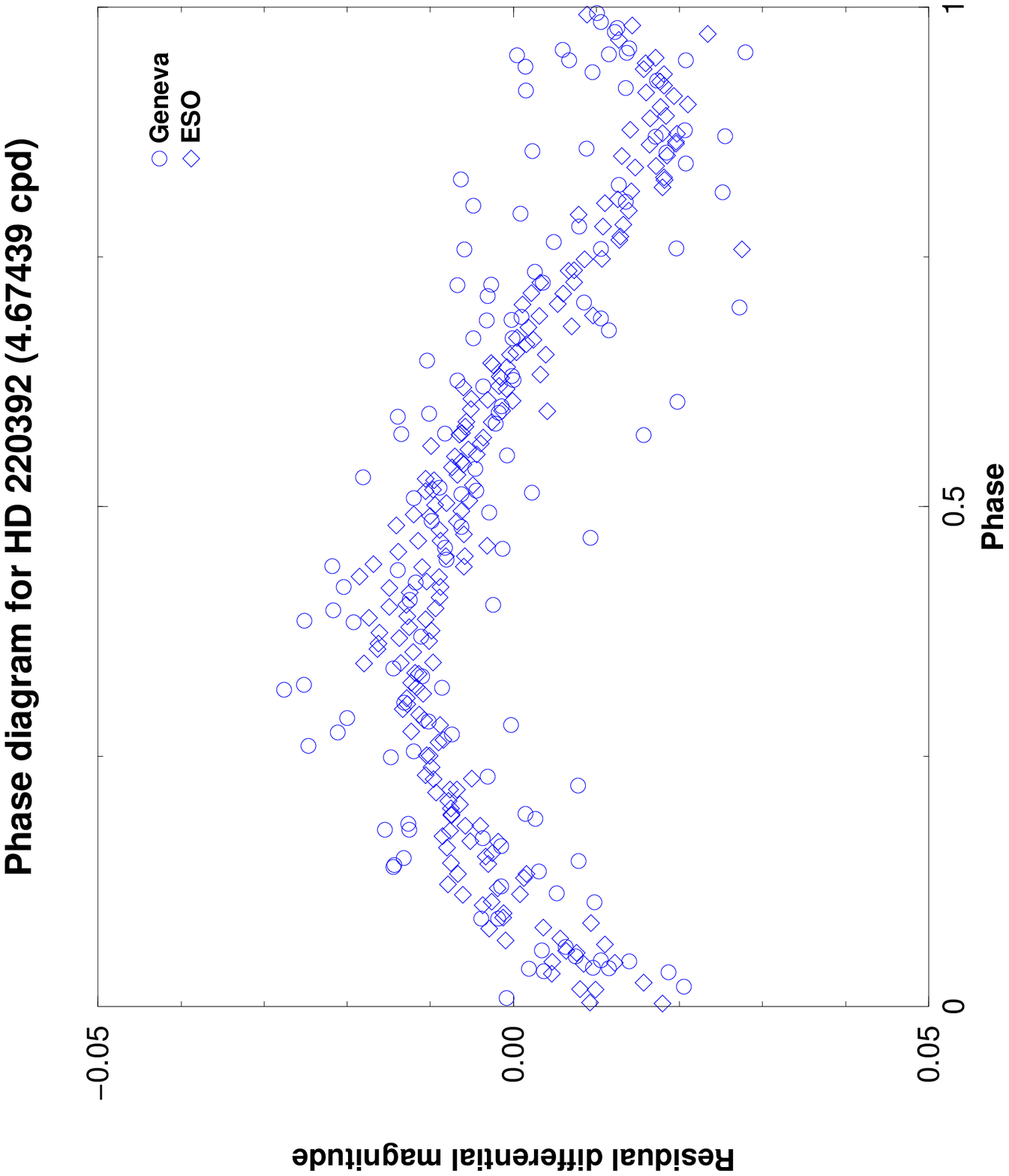}{0.0cm}{270}{30}{30}{-180}{25}
\plotfiddle{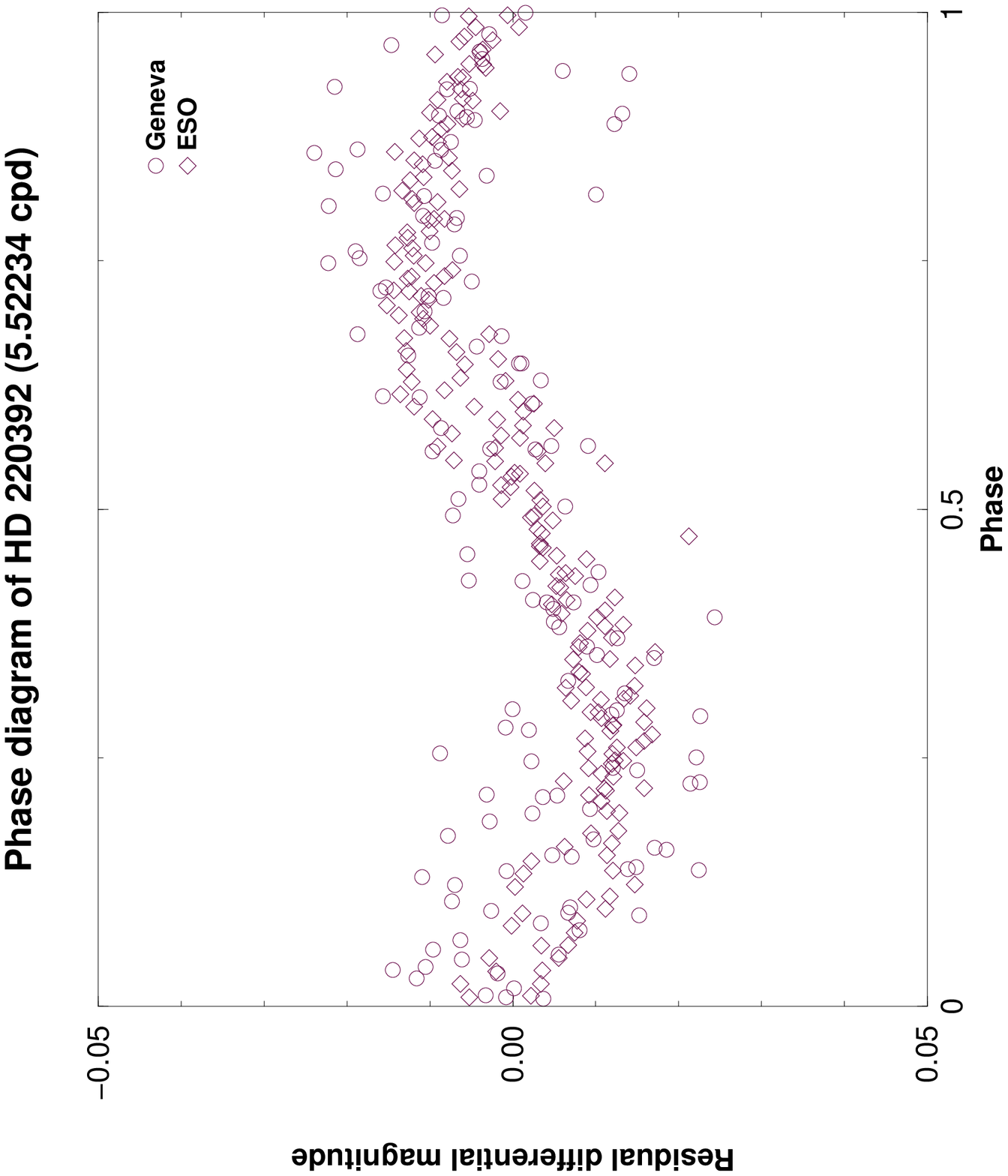}{-4.0cm}{270}{30}{30}{0}{50} 
\vspace{4cm}
\caption{Phase diagrams for the HD~220392-data against the frequencies of 4.67439 cpd 
(after removal for the 5.52 cpd variation) ({\it left}) and of 5.52234 cpd (after 
 removal for the 4.67 cpd variation) ({\it right}) 
\label{pha1122} }
\end{figure}

The Hipparcos Epoch Photometry Catalogue lists 183 measurements of HD~220392 (HIP~115510). 
The note in the Main Catalogue however mentions that the ``data are inadequate
for confirmation of the period from Ref. 94.191'' (ESA~1997). The reason for 
this are the quality flags that all are equal to or larger than 16, meaning ``possibly 
interfering object in either field of view''. The effective width
of the aperture is 38 arcsec, so companions at angular separations between 10 and 
30 arcsec may interfere significantly during the measurement. We rejected some 
suspicious data (Lampens et al.~1999). Fourier analysis of the remaining 176 data 
then revealed 4.6743 $\pm$ 0.0001 cpd, the same frequency as found in all former datasets. 
The amplitude associated with $f_{1}$ is 0.013 mag large but the second frequency  
(5.52 or 6.52 ~cpd) remained below detection as a two-frequency fit attributed 
an amplitude of only 0.003 mag to $f_{2}$.

\subsection{HD~220391}

Data were obtained during the last two seasons only. The standard
deviation of the 245 measurements is less than half the one of the
previously discussed dataset. A frequency search was performed in a
similar way as for HD~220392: one peak at the frequency of 0.42 cpd was
found but the associated amplitude of 0.005 mag is below the expected
noise level and the reduction of the standard deviation is too small.

The Hipparcos Epoch Photometry catalogue lists 182 measurements of
HD~220391 (HIP~115506).  Again all quality flags are equal to or larger
than 16. We selected 172 data by applying the same conservative criteria
as above. Fourier analysis between 0. and 23. cpd displayed a peak at
$\sim$ 11 cpd (with an associated amplitude of 0.013 mag!).  This artifact
frequency of order 2 hr$^{-1}$ is introduced by the rotation period of
the satellite.

\section{Astrophysical considerations}

\subsection{The nature of the association}

\begin{figure}[]
\plotfiddle{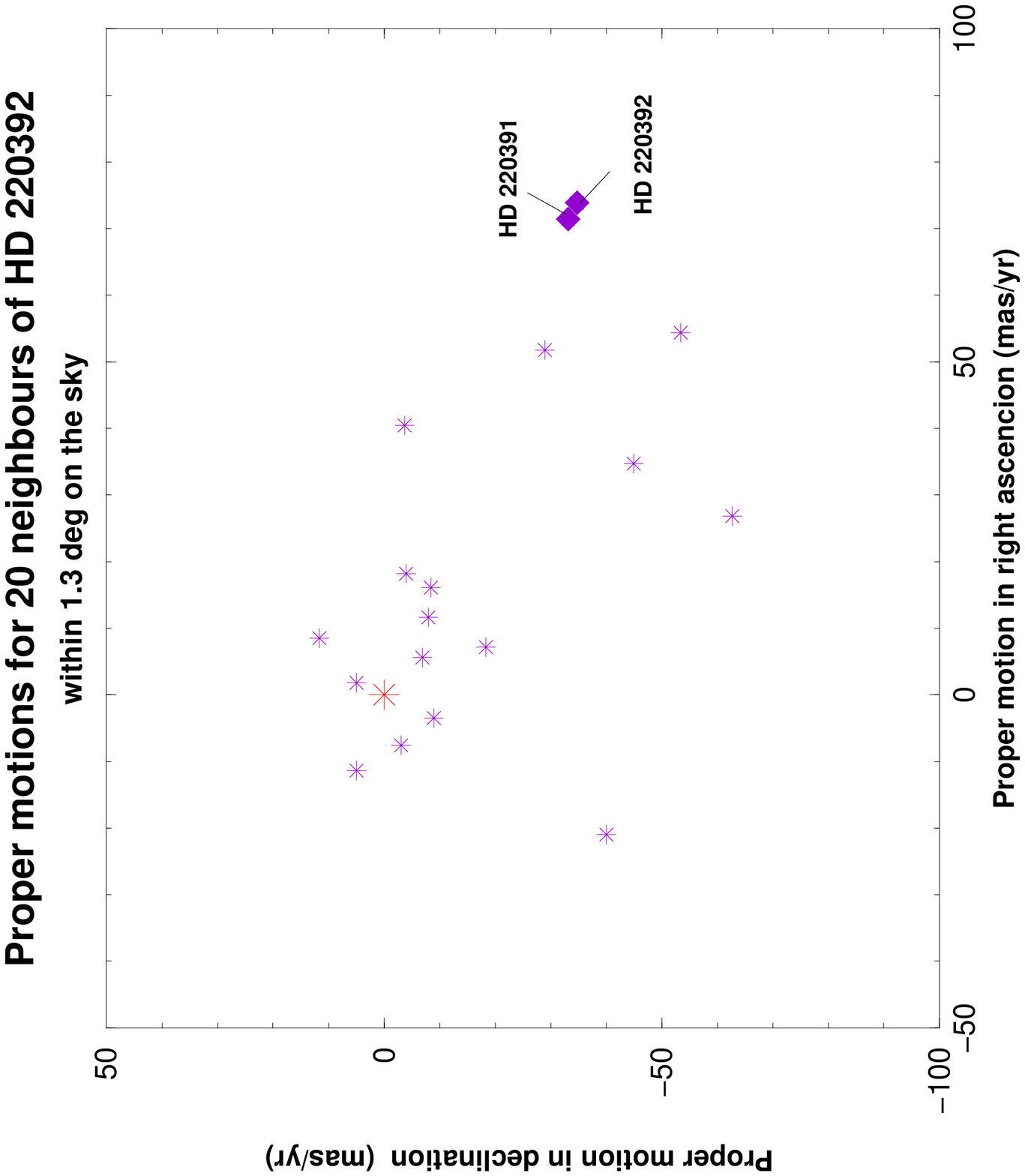}{0.0cm}{270}{30}{30}{-180}{25}
\plotfiddle{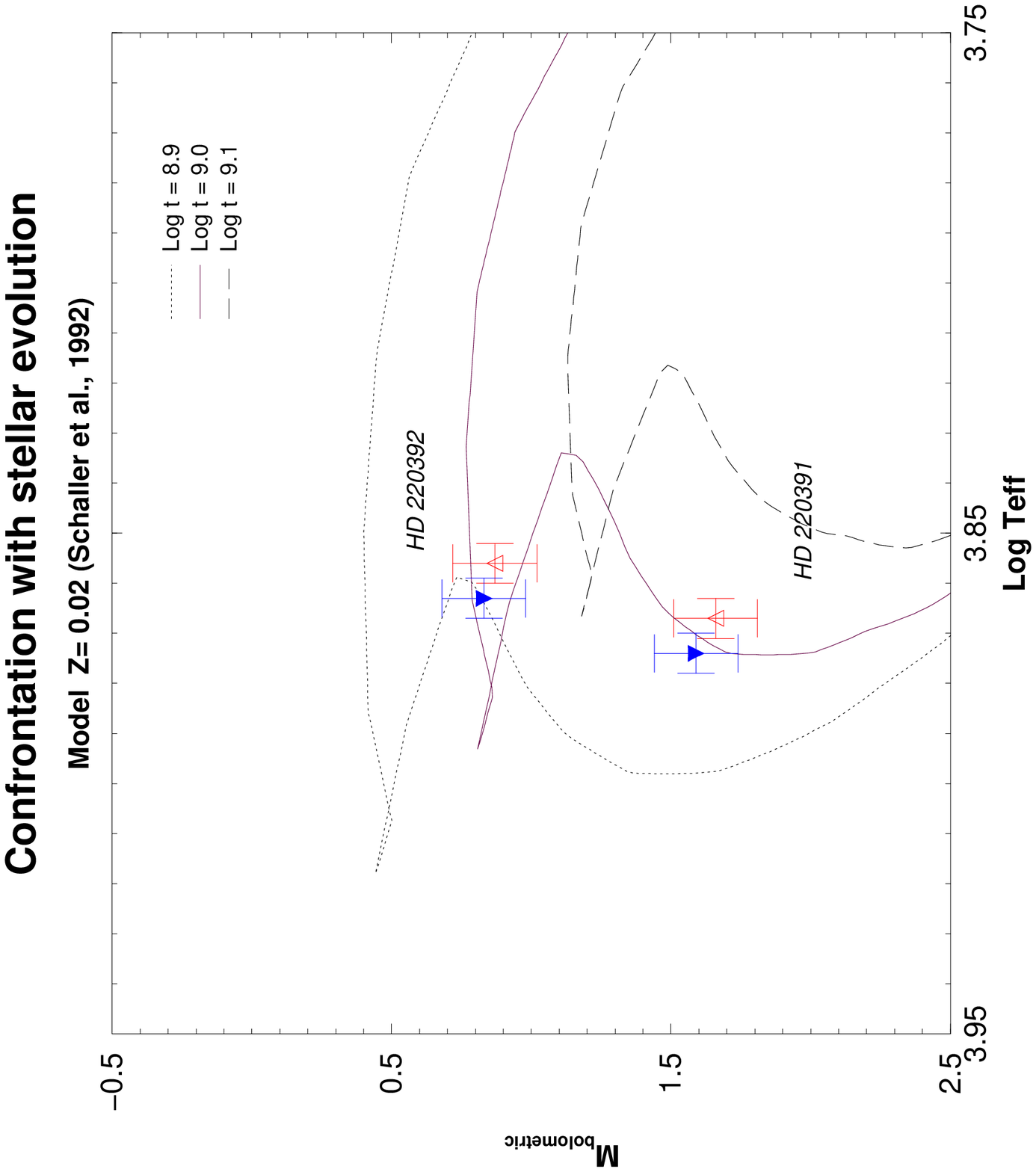}{-4.0cm}{270}{30}{30}{0}{50} 
\vspace{4cm}
\caption{Proper motion distribution in the sky area around HD 220392/1 ({\it left})
 and isochrone fit in the HR diagram (Z=0.020) ({\it right}). Filled symbols refer to the locations
 after the correction for rotation.  
\label{pmiso} }
\end{figure}

From the mean Geneva colour indices and the corresponding calibrations
for A-F type stars in the Geneva Photometric System (K\"{u}nzli et
al.~1997; Kobi \& North 1990) we derived the physical parameters
presented in Table~\ref{phys}.  The Hipparcos astrometric data are
useful to establish the nature of the association: the relative proper
motion between the two components of this wide system is quite small
($\Delta \mu _{\alpha _{B-A}}\simeq -2.44$~milli-arcsec/yr (mas/yr),
$\Delta \mu _{\delta _{B-A}}\simeq +1.57$~mas/yr with errors of the same
order) while the parallaxes are compatible to better than $1.5\,\sigma $
($\pi _{A}=6.79\pm 1.43$ mas, $\pi _{B}=9.19\pm 2.44$ mas). In the left
part of Figure~\ref{pmiso}, all stars within 1.3\deg~ on the sky with
proper motions from the Simbad database have been plotted to illustrate
the concordance of the proper motions for both stars. These data confirm
the common proper motion status and the probable physical association
of the pair (van de Kamp~1982).  Radial velocities would be very useful
but such information is lacking for component B.  Both components also
share an identical {\it projected} rotational velocity.

We used the absolute magnitudes to fit a model of stellar evolution of
solar chemical composition (Schaller et al.~1992) in a theoretical H-R
diagram. The same isochrone with an estimated age of $\approx$ $10^{9}$
years for the system appears to fit both stars well (right part of
Figure~\ref{pmiso}). This conclusion holds even after removal of the
effects of rotation at the rate of half the break-up velocity (P\'erez
Hern\'andez et al.~1999), as illustrated by the filled symbols in the
same figure. We conclude that both stars thus form a common origin pair
and probably even a true binary system.

\subsection{The nature of the variability}

The mean (d,~B$_{2}$-V$_{1}$)-values place both stars well within the
$\delta $ Scuti instability strip as observed in the Geneva Photometric
System. We note the interesting situation where two physically associated
stars with similar characteristics behave quite differently from the
variability point-of-view.  In the previous sections we have shown that
the brightest component behaves as a $\delta$ Scuti variable with a total
amplitude of 0.05 mag while the fainter component presents no short-period
variability of amplitude larger than 0.01 mag. What could the causes be
for the difference in variability between both? From the Geneva colour
indices, it appears that the brightest component has $\Delta d>0.100$,
thus it is more evolved than its companion. From the isochrone fit, one
may also notice the probable core hydrogen burning phase of HD~220391 and
the overall contraction or shell hydrogen burning phase of the brightest
component, HD~220392. Evolution appears in this case to be the probable
cause for the observed diversity in variability.

\begin{table}[]
\caption{Physical parameters derived for HD~220392/1
\label{phys} }
\begin{center}
{
\small
\begin{tabular}{ccccc}
\hline
Identifiers & HD & 220392 & 220391 & Source\struutup \\ 
 &  Hip & 115510 & 115506 &\\ 
 & CCDM & 23239-5349A & 23239-5349B &\struutdown \\ \hline
Sp. Type &  & F0IVn & A9Vn & GG89 \struutup\\ 
${\rm m}_{v}$ & mag & 6.124 $\pm $ 0.014 & 7.103 $\pm $ 0.007 & \\ 
U & mag & 1.608 & 1.538 & \\ 
V & mag & 0.647 & 0.662 & G\\ 
B$_{1}$ & mag & 0.958 & 0.954 & E\\ 
B$_{2}$ & mag & 1.422 & 1.426 & N\\ 
V$_{1}$ & mag & 1.364 & 1.379 & E\\ 
G & mag & 1.769 & 1.790 & V\\ 
d & mag & 1.314 & 1.259 & A\\ 
B$_{2}-$V$_{1}$ & mag & 0.058 & 0.047 & \\ 
${\rm M}_{\rm V}$ & mag & +0.83 $\pm $ 0.15 & +1.62 $\pm $ 0.15 & \\ 
${\rm M}_{\rm bol}$ & mag & +0.87 $\pm $ 0.15 & +1.66 $\pm $ 0.15 & FL96 \\ 
log$T_{\rm eff}$ & K & 3.856 $\pm $ 0.008 & 3.867 $\pm $ 0.009 & \\ 
$\lbrack M/H]$ & dex & -0.05 $\pm $ 0.09 & -0.12 $\pm $ 0.10 & \\ 
log g & dex & 3.77 $\pm $ 0.07 & 4.06 $\pm $ 0.07 & \\ 
$ M $ & $ M_{\odot }$ & 2.3 $\pm $ 0.2 & 1.8 $\pm $ 0.2 & NO96 \struutdown\\ \hline
${\rm vsin}{\it i}$ & km$s^{-1}$ & 165 & 140 & LE75 \struutup\\ 
$\pi _{phot}$ & mas & 8.7 $\pm $ 1 & 8.0 $\pm $ 1 &  \\ 
$\pi _{Hip}$ & mas &  6.79 $\pm $ 1.43 & 9.19 $\pm $ 2.44& E \\ 
$\mu _{\alpha }^{\ast }$ & ''/yr & 0.073 & 0.071 & S \\ 
$\mu _{\delta }$ & ''/yr & -0.035 & -0.033 \struutdown & A \\ \hline
\end{tabular}
\\[0pt]
}
\end{center}
\end{table}

From the properties listed in Table~\ref{phys}, the pulsation constants can be 
computed but there is no clear conclusion at present: one of the frequencies ($f_{2}$) 
may possibly correspond to the fundamental radial mode (F)(Q = 0.037 or 0.032 days). 
Additional photometric observations for this interesting couple of stars is 
certainly recommended. The already obtained data are neither sufficiently numerous 
nor of sufficient quality to allow unambiguous solutions or to solve for the multiple 
frequencies. Radial velocities would be needed too. 

\section{Conclusion}

What factors actually determine the pulsation characteristics,
i.e., the amplitudes and modes of pulsation in the $\delta$ Scuti
instability strip? This question cannot be addressed on the basis
of a single case. But it could be approached as illustrated here.
Binary systems with pulsating variable components offer a unique
opportunity of coupling the information obtained by astrometric means
(association type - parallax - total mass) to the astrophysical quantities
(luminosity ratio - colours - pulsation characteristics). For example,
the detailed investigation of differences in variability between the
components of a binary may provide relevant clues with respect to
their pulsation characteristics. Differences in origin and age can
be ruled out as well as differences in overall chemical composition.
In this case the short-period pulsating component is easily identified
and the information obtained on the variability can be coupled to the
astrophysical parameters of {\it each} component. It would be even better 
to investigate such characteristics in a visual binary for which basic
information on the orbital motion can be derived.  This would allow one to
obtain a direct estimation of the stellar mass, independent of any choice
of modelisation. The derivation of the pulsation constant would be more
straightforward too (the error on the mass defines the accuracy of Q).

\acknowledgments
We thank the Geneva team (G.~Burki) for the telescope time put at our 
disposal in June~1990 and September~1991 as well as D.~Sinachopoulos for the 
multiple observations performed at the ESO 0.5m telescope. M.~Sperl is kindly acknowledged 
for making the programme Period98 available for this application. We appreciate
the help of L. Eyer in the selection of the Hipparcos Epoch Photometry data. We
made use of the Simbad database operated at CDS, Strasbourg, France.

\end{document}